\documentclass[prd,aps,nofootinbib,preprint]{revtex4}
\usepackage{graphicx,epsfig}

\def\pp{{\prime\prime}}
\def\vp{\varepsilon}

\begin{document}

\title{The study of $B_c^- \to X(3872)\pi^-(K^-)$ Decays in the
Covariant Light-Front Approach }
\author{Wei Wang, Yue-Long Shen and Cai-Dian L\"u }
\affiliation{Institute of High Energy Physics, Chinese Academy of
Sciences, Beijing 100049, P.R. China }

\begin{abstract}
In the covariant light-front quark model, we  calculate the form
factors of $B_c^-\to J/\psi$ and $B_c^-\to X(3872)$. Since the
factorization of the exclusive processes $B_c^- \to
J/\psi\pi^-(K^-)$ and $B_c^- \to X(3872)\pi^-(K^-)$ can be proved
in the soft-collinear effective theory, we can get the branching
ratios for these decays easily from the form factors.
 Taking the uncertainties into account, our results
for the branching ratio of $B_c^-\to J/\psi \pi^-(K^-)$ are
consistent with the previous studies. By identifying $X(3872)$  as a
$1^{++}$ charmonium state, we obtain ${\cal BR}(B_c^-\to
X(3872)\pi^-)=(1.7^{+0.7+0.1+0.4}_{-0.6-0.2-0.4}) \times 10^{-4}$
and ${\cal BR}(B_c^-\to
X(3872)K^-)=(1.3^{+0.5+0.1+0.3}_{-0.5-0.2-0.3})\times 10^{-5}$. If
assuming $X(3872)$ as a $1^{--}$ state, the branching ratios will be
one order magnitude larger than those of $1^{++}$ state. These
results can be easily used to test the charmonium description for
this mysterious meson $X(3872)$ at LHCb experiment.
\end{abstract}
\maketitle
\newpage


\section{Introduction}

$X(3872)$ was first observed by Belle in the exclusive decay
$B^{\pm}\rightarrow K^{\pm}X\rightarrow K^{\pm}\pi^+\pi^- J/\psi$
\cite{BelleX3872}, and subsequently confirmed by CDF, D0 and BaBar
collaborations in various decay and production channels
\cite{X3872other}. At present a definite answer on its internal
properties is not well established, but the current experimental
data strongly support a $1^{++}$ state \cite{X3872JPC}. Enormous
interest in $\bar cc$ resonance spectroscopy study followed this
discovery and there exists many interpretations for this meson. The
first and most natural assignment of this state is the first radial
excitation of $1P$ charmonium state $\chi_{c1}$~\cite{charmoniumBG}.
However, this interpretation has encountered two difficulties: its
decay width($<2.3$ MeV, 95$\%$ C.L.) is tiny compared with other
charmonia; there is a gap of about 100 MeV between the measured mass
and the quark model prediction~\cite{reviewX}. Motivated by these
two difficulties, many non-charmonium explanations  were proposed,
such as $\bar ccg$ hybrid meson \cite{hybrid},
glueball~\cite{glueball}, diquark cluster \cite{tetra}, and
molecular state \cite{molecule}. But in fact, there is few
experimental data which could provide a clear discrimination among
these descriptions and this makes the situation more obscure.
Recently the CLQCD collaboration studied the mass for the first
excited states of $1^{++}$ charmonium and found it is consistent
with the measured mass of $X(3872)$~\cite{CLQCDX}. The consistence
indicates that $X(3872)$ can be the first radial excited state of
$\chi_{c1}$ and it seems that the mass difficulty trails off. Now,
in order to investigate the structure of this meson more clearly, a
large amount of experimental data and theoretical studies on the
productions and decays of $X(3872)$ are strongly deserved.

In $B_{u,d,s}$ decays involving charmonium final sates, the
emitted meson is a heavy charmonium. The non-factorizable
contribution should be large to induce large uncertainties
\cite{Cheng:2001ez}. As the energy release is limited, these
decays may also be  polluted by the final state interactions which
are non-perturbative in nature.  But fortunately the production of
charmonia in $B_c$ decays could provide a unique insight to these
mesons. Since the emitted meson here is a light meson ($\pi$ or
$K$), the factorization of $B_c\to(\bar cc) M$ (M is a light
meson) could be proved in the framework of
Soft-Collinear-Effective-Theory (SCET) to all orders of the strong
coupling constant in heavy quark limit which is similar with $\bar
B^0\to D^+\pi^-$ and $B^-\to D^0\pi^-$~\cite{SCETDpi}. The decay
matrix element can be decomposed into the $B_c\to (\bar cc)$ form
factor and a convolution of a short distance coefficient with the
light-cone wave function of the emitted light meson.

Although SCET provides a powerful framework to study the
factorization of the exclusive modes, the non-perturbative form
factors could not be directly studied. We can only  extract them
via the experimental data or rely on some non-perturbative method.
In the present paper, we will use the light-front quark model to
calculate these $B_c\to M(\bar cc)$ form factors. As pointed out
in \cite{BPP}, the light front QCD approach has some unique
features which are particularly suitable to describe a hadronic
bound state.  The light-front quark model~\cite{Jaus1,CCH1,CJK}
can provide a relativistic treatment of the movement of the hadron
and also give a fully treatment of the hadron spin by using the
so-called Melosh rotation.  Light-front wave functions, which
describe the hadron in terms of their fundamental quark and gluon
degrees of freedom, are independent of the hadron momentum and
thus are explicitly Lorentz invariant. Furthermore, in covariant
light-front approach\cite{Jaus2}, the spurious contribution which
is dependent on the orientation of the light-front is eliminated
by including the zero-mode contributions properly. This covariant
model has been successfully extended to study the decay constants
and form factors of the ground-state $s$-wave and low-lying
$p$-wave mesons~\cite{CCH2,Cheng:2004yj,LFHW}.

The paper is organized as follows. The formalism for the form factor
calculations,   taking $B_c^-\to J/\psi$ as an example, is presented
in the next section.  The numerical results for form factors and
decay rates of $B_c^-\to J/\psi\pi^-(K^-)$, $B_c^-\to
J/\psi\rho^-(K^{*-})$, $B_c^-\to X(3872)\pi^-(K^-)$ and $B_c^-\to
X(3872)\rho^-(K^{*-})$ are given in Section \ref{III}.  The
conclusion is given in Section~\ref{conc}.

\section{Calculation of the form factors and the branching ratios}

In the following, we use $X$ to denote $X(3872)$ for simplicity.
Different with $B_{u,d,s}$ mesons, the $B^-_c$ system consists of
two heavy quarks $b$ and $c$, which can decay individually. Here
we will consider $b$ decays while $\bar c$ acts as a spectator. At
the quark level, $B_c^-\to J/\psi\pi^-$ and $ B_c^-\to
X(3872)\pi^-$ decays are characterized by $b\to (cd\bar u)$
transition and the corresponding effective Hamiltonian  is given
by~\cite{Buras} :
 \begin{eqnarray}
 {\cal H}_{eff} &=& \frac{G_{F}}{\sqrt{2}}
     V_{cb} V_{ud}^{\ast} \Big[
     C_{1}(\mu)O_{1}(\mu)
  +  C_{2}(\mu) O_{2}(\mu)\Big] + \mbox{H.c.},
 \label{eq:hamiltonian01}
 \end{eqnarray}
where $V_{ij}$ are the corresponding CKM matrix elements. The
local four-quark operators $O_{1,2}$ are defined by:
\begin{eqnarray}
  O_{1}(\mu)=({\bar{c}}_{\alpha}b_{\beta} )_{V-A}
               ({\bar{d}}_{\beta} u_{\alpha})_{V-A},
    \ \ \ \ \ \ \ \ \
   O_{2}(\mu)=({\bar{c}}_{\alpha}b_{\alpha})_{V-A}
               ({\bar{d}}_{\beta} u_{\beta} )_{V-A},
    \end{eqnarray}
where $\alpha$ and $\beta$ are the color indices. Since the four
quarks in the operators are different with each other, there is no
penguin contribution and thus there is no CP violation. The left
handed current is defined as $({\bar{q}}_{\alpha} q_{\beta}
)_{V-A}= {\bar{q}}_{\alpha} \gamma_\nu (1-\gamma_5) q_{\beta}  $.
For $b\to (cs\bar u)$ transition, $V_{ud}^*$ is replaced by
$V_{us}^*$ while $d$ quark field in the four-quark operator is
replaced by $s$. With the effective Hamiltonian given above, the
matrix element for the $B_c^-\to J/\psi\pi^-$ transition can be
expressed as:
 \begin{eqnarray}
{\cal M}=\langle J/\psi(P^\pp,\vp^{\pp*})\pi|{\cal H}_{ eff}|
 B_{c}^-(P^\prime)\rangle = \frac{G_{F}}{\sqrt{2}}
     V_{cb} V_{ud}^{\ast} a_1(\mu) \langle J/\psi(P^\pp,\vp^{\pp*})\pi|
      O_2(\mu)|
 B_{c}^-(P^\prime)\rangle  ,
 \end{eqnarray}
with $P^{\prime(\pp)}$ being the incoming (outgoing) momentum,
$\vp^{\pp*}$ the polarization vector of $J/\psi$ and
$a_{1}=C_2+C_1/3$ the Wilson coefficient.

In the effective Hamiltonian, the degrees of freedom heavier than
$b$ quark mass $m_b$ scale are included in the Wilson coefficients
which can be calculated using the perturbation theory. Then the
left task is to calculate the operators' matrix elements between
the $B_c^-$ meson state and the final states, which suffers large
uncertainties. Nevertheless, the problem becomes tractable if
factorization becomes applicable. Thanks to the development of
SCET, the proof of the factorization can be accomplished in an
elegant way~\cite{SCET1,SCET2}. In SCET, the heavy meson is
described by the heavy quarks $h_v$ and soft gluons $A_s$ in its
rest frame; the final state light meson moves very fast and it is
described by the collinear quarks $\xi_c$ and collinear gluons
$A_c$. In Ref.~\cite{SCETDpi}, it has been shown that the
collinear gluons do not connect to the particle in the heavy meson
while the soft gluons don't connect to those in the light meson to
all orders in $\alpha_s$ and leading power in
$\Lambda_{QCD}/m_{B_c}$. In a phenomenal language,  the
non-factorizable diagrams cancel with each other because of color
transparency. Furthermore, there is no annihilation contribution
as the quarks in the final state meson are different with each
other. Thus the decay amplitude can be expressed as the product of
$B_c\to J/\psi$ form factor and a convolution of a short distance
Wilson coefficient with non-perturbative light-cone distribution
amplitude of the light meson. Without the higher order QCD
corrections, the convolution is reduced to the decay constant of
the light meson.

The form factors for $B_c\to J/\psi$ and $B_c\to X(3872)$
($1^{++}$ state) transitions induced by the vector and
axial-vector currents are defined by:
{\small
\begin{eqnarray}
      \langle J/\psi(P^\pp,\vp^{\pp*})|V_\mu|B_c^-(P^\prime)\rangle &=&
       -\frac{1}{ m_{B_c}+m_{J/\psi}}\epsilon_{\mu\nu\alpha \beta}\vp^{\pp*\nu}P^\alpha q^\beta  V^{PV}(q^2),    \\
\ \ \ \
      \langle J/\psi(P^\pp,\vp^{\pp*})|A_\mu|B_{c}^-(P^\prime)\rangle &=& i\Big\{
         (m_{B_c}+m_{J/\psi})\vp^{\pp*}_\mu A_1^{PV}(q^2)-\frac{\vp^{\pp*}\cdot P}
         { m_{B_c}+m_{J/\psi}}
         P_\mu A_2^{PV}(q^2)    \nonumber \\
    && -2m_{J/\psi}\,{\vp^{\pp*}\cdot P\over
    q^2}q_\mu\big[A_3^{PV}(q^2)-A_0^{PV}(q^2)\big]\Big\},\\
\ \ \ \
       \langle X(P^\pp,\vp^\pp)|V_\mu|B_c^-(P^\prime)\rangle &=&
        (m_{B_c}-m_X) \vp^*_\mu V_1^{PA}(q^2)-\frac{\vp^*\cdot
      P'}{ m_{B_c}-m_X}P_\mu V_2^{PA}(q^2) \nonumber \\
      &&- 2m_X {\vp^*\cdot P'\over
         q^2}q_\mu\left[V_3^{PA}(q^2)-V_0^{PA}(q^2)\right], \\
\ \ \
      \langle X(P^\pp,\vp^\pp)|A_\mu|
          B_{c}^-(P^\prime)\rangle &=& -\frac{i}{
          m_{B_c}-m_X}\epsilon_{\mu\nu\rho\sigma}\vp^{*\nu}P^\rho
       q^{\sigma}A^{PA}(q^2).
 \end{eqnarray}}
where $P=P^\prime+P^{\prime\prime}$, $q=P^\prime-P^{\prime\prime}$
and the convention $\epsilon_{0123}=1$ is adopted. To cancel the
poles at $q^2=0$, we must have $A_3^{PV}(0)=A_0^{PV}(0)$,
$V_3^{PA}(0)=V_0^{PA}(0)$. The form factor $A_3^{PV}(V_3^{PA})$ is
related to other form factors by:
\begin{eqnarray}
A_3^{PV}(q^2)&=&\frac{m_{B_c}+m_{J/\psi}}{ 2m_{J/\psi}}
A_1^{PV}(q^2)-\frac{m_{B_c}-m_{J/\psi}}{ 2m_{J/\psi}}\,A_2^{PV}(q^2), \nonumber\\
V_3^{PA}(q^2)&=&\frac{m_{B_c}-m_X}{
2m_X}\,V_1^{PA}(q^2)-\frac{m_{B_c}+m_X}{ 2m_X}\,V_2^{PA}(q^2).
 \end{eqnarray}

\begin{figure}
\centerline{{\epsfxsize 3 in
 \epsffile{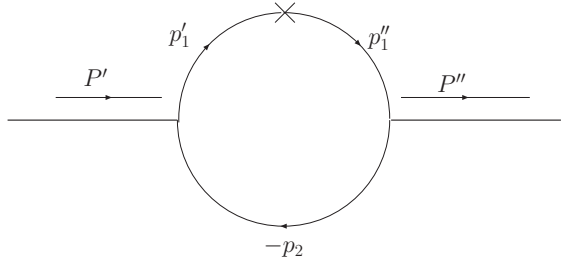}}}
\caption{Feynman diagram for $B_c\to J/\psi(X(3872))$ decay
amplitudes. The $X$ in the diagram denotes the $V-A$ transition
vertex while the meson-quark-antiquark vertices are given in the
text.}\label{fig:feyn}
\end{figure}

Following the notation in Refs.~\cite{Jaus2,CCH2}, we use the
light-front decomposition of the momentum $P^{\prime}=(P^{\prime
-}, P^{\prime +}, P^\prime_\bot)$, where
$P^{\prime\pm}=P^{\prime0}\pm P^{\prime3}$, so that $P^{\prime
2}=P^{\prime +}P^{\prime -}-P^{\prime 2}_\bot$ and work in the
$q^+=0$ frame. The incoming and outgoing mesons have the momentum
$P^{\prime}=p_1^{\prime}+p_2$ and $P^{\pp}=p_1^{\pp}+p_2$,
respectively.  The quark and antiquark inside the incoming
(outgoing) meson have the mass $m_1^{\prime(\pp)}$ and $m_2$ whose
momenta are denoted as $p_1^{\prime(\pp)}$ and $p_2$ respectively.
These momenta can be expressed in terms of the internal variables
$(x_i, p_\bot^\prime)$ as:
 \begin{eqnarray}
 p_{1,2}^{\prime+}=x_{1,2} P^{\prime +},\qquad
 p^\prime_{1,2\bot}=x_{1,2} P^\prime_\bot\pm p^\prime_\bot,
 \end{eqnarray}
with $x_1+x_2=1$. Using these internal variables, we can define
some other useful quantities for the incoming meson:
\begin{eqnarray}
 M^{\prime2}_0
          &=&(e^\prime_1+e_2)^2=\frac{p^{\prime2}_\bot+m_1^{\prime2}}
                {x_1}+\frac{p^{\prime2}_{\bot}+m_2^2}{x_2},\quad\quad
                \widetilde M^\prime_0=\sqrt{M_0^{\prime2}-(m^\prime_1-m_2)^2},
 \nonumber\\
 e^{(\prime)}_i
          &=&\sqrt{m^{(\prime)2}_i+p^{\prime2}_\bot+p^{\prime2}_z},\quad\qquad
 p^\prime_z=\frac{x_2 M^\prime_0}{2}-\frac{m_2^2+p^{\prime2}_\bot}{2 x_2 M^\prime_0}.
 \end{eqnarray}
$e_i^{(\prime)}$ can be interpreted as the energy of the quark or
the antiquark and $M_0^\prime$ can be viewed kinematic invariant
mass of the meson system. To calculate the amplitude for the
transition form factor, we need the following Feynman rules for
the meson-quark-antiquark vertices
($i\Gamma^\prime_M$)\footnote{We use a different phase $-i$ with
Ref.~\cite{CCH2} for the incoming axial-vector vertex.}:
\begin{eqnarray}
i\Gamma_P^\prime&=&
      H^\prime_P\gamma_5,
            \\
i\Gamma_V^\prime&=&iH^\prime_{V}[\gamma_\mu-\frac{1}{W^\prime_{V}}(p^\prime_1-p_2)_\mu],
            \\
i\Gamma_A^\prime&=&-H^\prime_{A}[\gamma_\mu+\frac{1}{W^\prime_{A}}(p^\prime_1-p_2)_\mu]
\gamma_5.
\end{eqnarray}
Here and in the following, we use the subscript $A$ to denote the
axial-vector with the quantum numbers $J^{PC}=1^{++}$.  For the
outgoing meson, we should use $i(\gamma_0
\Gamma^{\prime\dagger}_M\gamma_0)$ for the corresponding vertices.

In the conventional light-front quark model, the constituent
quarks are required to be on mass shell and the physical
quantities can be extracted from the plus component of the
corresponding current matrix elements. However, this framework
suffers the problem of non-covariance and miss the zero-mode
contributions. To solve this problem, Jaus proposed the covariant
light-front approach which can deal with the zero mode
contributions systematically~\cite{Jaus2}. Decay constants and
form factors can be calculated in terms of Feynman momentum loop
integrals which are manifestly covariant. In this framework, the
lowest order contribution to a form factor is depicted in
Fig.~\ref{fig:feyn}. For the $P\to V$ transition, the decay
amplitudes are:
 \begin{eqnarray}
 {\cal B}^{PV}_\mu=-i^3\frac{N_c}{(2\pi)^4}\int d^4 p^\prime_1
 \frac{H^\prime_P (i H^\pp_V)}{N_1^\prime N_1^\pp N_2} S^{PV}_{\mu\nu}\,\vp^{\pp*\nu},
 \end{eqnarray}
where
$N_1^{\prime(\prime\prime)}=p_1^{\prime(\prime\prime)2}-m_1^{\prime(\prime\prime)2}+i\epsilon$,
$N_2=p_2^2-m_2^2+i\epsilon$ and{\small
 \begin{eqnarray}
S^{PV}_{\mu\nu} &=&(S^{PV}_{V}-S^{PV}_A)_{\mu\nu}
 \nonumber\\
                &=&{\rm Tr}\left[\left(\gamma_\nu-\frac{1}{W^\pp_V}
                (p_1^\pp-p_2)_\nu\right)
                                 (\not \!p^\pp_1+m_1^\pp)
                                 (\gamma_\mu-\gamma_\mu\gamma_5)
                                 (\not \!p^\prime_1+m_1^\prime)\gamma_5(-\not
                                 \!p_2+m_2)\right]\nonumber \\
           &=&-2 i\epsilon_{\mu\nu\alpha\beta}
                 \Big\{p^{\prime\alpha}_1 P^\beta (m_1^\pp-m_1^\prime)
                   +p^{\prime\alpha}_1q^\beta(m_1^\pp+m_1^\prime-2
                   m_2)+q^\alpha P^\beta m_1^\prime
                 \Big\}
 \nonumber\\
           &&+\frac{1}{W^\pp_V}(4 p^\prime_{1\nu}-3
           q_\nu-P_\nu)i\epsilon_{\mu\alpha\beta\rho}p^{\prime\alpha}_1 q^\beta P^\rho
 \nonumber\\
           &&+2
           g_{\mu\nu}\Big\{m_2(q^2-N_1^\prime-N^\pp_1-m_1^{\prime2}-m_1^{\pp2})
                      -m_1^\prime (M^{\pp2}-N_1^\pp-N_2-m_1^{\pp2}-m_2^2)
 \nonumber\\
           &&-m^\pp_1(M^{\prime2}-N^\prime_1-N_2-m_1^{\prime2}-m_2^2)
           -2 m_1^\prime m_1^\pp m_2\Big\}
  \nonumber\\
           &&+8 p^\prime_{1\mu} p^\prime_{1\nu}(m_2-m_1^\prime)
             -2(P_\mu q_\nu+q_\mu P_\nu+2q_\mu q_\nu) m_1^\prime
             +2p^\prime_{1\mu} P_\nu (m_1^\prime-m_1^\pp)
 \nonumber\\
           &&+2 p^\prime_{1\mu} q_\nu(3 m_1^\prime-m_1^\pp-2m_2)+2
           P_\mu p^\prime_{1\nu}(m_1^\prime+m_1^\pp)+2 q_\mu
           p^\prime_{1\nu}(3 m_1^\prime+m_1^\pp-2 m_2)
 \nonumber\\
           &&+\frac{1}{2W^\pp_V}(4 p^\prime_{1\nu}-3q_\nu-P_\nu)
              \Big\{2 p^\prime_{1\mu}[M^{\prime2}+M^{\pp2}-q^2
                -2 N_2+2(m_1^\prime-m_2)(m_1^\pp+m_2)]
 \nonumber\\
           &&   +q_\mu[q^2-2 M^{\prime2}+N^\prime_1-N_1^\pp+2 N_2-(m_1+m_1^\pp)^2
           +2(m_1^\prime-m_2)^2]
 \nonumber\\
           &&   +P_\mu[q^2-N_1^\prime-N_1^\pp-(m_1^\prime+m_1^\pp)^2]
              \Big\}.
              \label{eq:BtoV}
 \end{eqnarray}}

In practice, we use the light-front decomposition of the loop
momentum and have to perform the integration over the minus
component using the contour method, as the covariant vertex
functions can not be determined by solving the bound state
equation. If the covariant vertex functions are not singular when
performing the integration, the transition amplitude will pick up
the singularities in the antiquark propagator.  The integration
leads to:
 \begin{eqnarray}
 N_1^{\prime(\pp)}  &\to&   \hat N_1^{\prime(\pp)}=x_1(M^{\prime(\pp)2}-M_0^{\prime(\pp)2}), \nonumber\\
 H^{\prime(\pp)}_M  &\to&   h^{\prime(\pp)}_M, \nonumber\\
 W^\pp_M            &\to&    w^\pp_M, \nonumber\\
\int \frac{d^4p_1^\prime}{N^\prime_1 N^\pp_1 N_2}H^\prime_P
 H^\pp_V S          &\to&  -i \pi \int \frac{d x_2d^2p^\prime_\bot}{x_2\hat N^\prime_1\hat N^\pp_1} h^\prime_P h^\pp_V \hat S,
 \end{eqnarray}
where
\begin{eqnarray}
 M^{\pp2}_0 =\frac{p^{\pp2}_\bot+m_1^{\pp2}}{x_1}+\frac{p^{\pp2}_{\bot}+m_2^2}{x_2},
 \end{eqnarray}
with $p^\pp_\bot=p^\prime_\bot-x_2\,q_\bot$. The explicit forms of
$h^\prime_M$ and $w^\prime_M$ for the pseudoscalar, vector and
axial-vector $1^{++}$ are given by~\cite{CCH2}
\begin{eqnarray}
 h^\prime_P&=&      h^\prime_V
                  =(M^{\prime2}-M_0^{\prime2})\sqrt{\frac{x_1 x_2}{N_c}}
                    \frac{1}{\sqrt{2}\widetilde M^\prime_0}\varphi^\prime,
 \nonumber\\
 h^\prime_{A}
                  &=&(M^{\prime2}-M_0^{\prime2})\sqrt{\frac{x_1 x_2}{N_c}}
                    \frac{1}{\sqrt{2}\widetilde M^\prime_0}\frac{\widetilde
                     M^{\prime
                     2}_0}{2\sqrt{2}M^\prime_0}\varphi^\prime_p,\\
                     w^\prime_V&=&M^\prime_0+m^\prime_1+m_2,\quad
 w^\prime_{A}=\frac{\widetilde{M}'^2_0}{m^\prime_1-m_2},\label{eqs:wa}
\end{eqnarray}
where $\varphi'$ and $\varphi'_p$ are the phenomenological
light-front momentum distribution amplitudes for $s$-wave and
$p$-wave mesons, respectively. After this integration, the
conventional light-front model is recovered but manifestly the
covariance is lost as it receives additional spurious
contributions proportional to the lightlike four vector
$\tilde\omega=(\tilde\omega^-,\tilde\omega^+,\tilde\omega_\perp)=(1,0,0_\perp)$.
The spurious contributions can be eliminated by including the zero
mode contribution which amounts to performing the $p^-$
integration in a proper way~\cite{Jaus2,CCH2}.

By using Eqs.~(\ref{eq:BtoV})--(\ref{eqs:wa}) and the integration
rules in Refs.~\cite{Jaus2,CCH2}, one arrives at{\small
\begin{eqnarray}
 g(q^2)&=&-\frac{N_c}{16\pi^3}\int dx_2 d^2 p^\prime_\bot
           \frac{2 h^\prime_P h^\pp_V}{x_2 \hat N^\prime_1 \hat N^\pp_1}
           \Bigg\{x_2 m_1^\prime+x_1 m_2+(m_1^\prime-m_1^\pp)
           \frac{p^\prime_\bot\cdot q_\bot}{q^2}
           +\frac{2}{w^\pp_V}\left[p^{\prime2}_\bot+\frac{(p^\prime_\bot\cdot
            q_\bot)^2}{q^2}\right]
           \Bigg\},
  \nonumber\\
  f(q^2)&=&\frac{N_c}{16\pi^3}\int dx_2 d^2 p^\prime_\bot
            \frac{ h^\prime_P h^\pp_V}{x_2 \hat N^\prime_1 \hat N^\pp_1}
            \Bigg\{2
            x_1(m_2-m_1^\prime)(M^{\prime2}_0+M^{\pp2}_0)-4 x_1
            m_1^\pp M^{\prime2}_0+2x_2 m_1^\prime q\cdot P
  \nonumber\\
         &&+2 m_2 q^2-2 x_1 m_2
           (M^{\prime2}+M^{\pp2})+2(m_1^\prime-m_2)(m_1^\prime+m_1^\pp)^2
           +8(m_1^\prime-m_2)\left[p^{\prime2}_\bot+\frac{(p^\prime_\bot\cdot
            q_\bot)^2}{q^2}\right]
  \nonumber\\
         &&
           +2(m_1^\prime+m_1^\pp)(q^2+q\cdot
           P)\frac{p^\prime_\bot\cdot q_\bot}{q^2}
           -4\frac{q^2 p^{\prime2}_\bot+(p^\prime_\bot\cdot q_\bot)^2}{q^2 w^\pp_V}
            \Bigg[2 x_1 (M^{\prime2}+M^{\prime2}_0)-q^2-q\cdot P
 \nonumber\\
         &&-2(q^2+q\cdot P)\frac{p^\prime_\bot\cdot
            q_\bot}{q^2}-2(m_1^\prime-m_1^\pp)(m_1^\prime-m_2)
            \Bigg]\Bigg\},
  \nonumber\\
a_+(q^2)&=&\frac{N_c}{16\pi^3}\int dx_2 d^2 p^\prime_\bot
            \frac{2 h^\prime_P h^\pp_V}{x_2 \hat N^\prime_1 \hat N^\pp_1}
            \Bigg\{(x_1-x_2)(x_2 m_1^\prime+x_1 m_2)-[2x_1
            m_2+m_1^\pp+(x_2-x_1)
            m_1^\prime]\frac{p^\prime_\bot\cdot q_\bot}{q^2}
  \nonumber\\
         &&-2\frac{x_2 q^2+p_\bot^\prime\cdot q_\bot}{x_2 q^2
            w^\pp_V}\Big[p^\prime_\bot\cdot p^\pp_\bot+(x_1 m_2+x_2 m_1^\prime)
            (x_1 m_2-x_2 m_1^\pp)\Big]
            \Bigg\},
 \end{eqnarray}}
while the physical form factors are related to the above functions
by
 \begin{eqnarray}
 V^{PV}(q^2)&=&-(m_{B_c}+m_{J/\psi})\, g(q^2),\quad
                 A_1^{PV}(q^2)=-\frac{f(q^2)}{m_{B_c}+m_{J/\psi}},
 \nonumber\\
 A_2^{PV}(q^2)&=&(m_{B_c}+m_{J/\psi})\, a_+(q^2).
 \end{eqnarray}

The extension to $P\to A$ transitions is straightforward:
 \begin{eqnarray}
 {\cal B}^{PA}_\mu&=&-i^3\frac{N_c}{(2\pi)^4}\int d^4 p^\prime_1
 \frac{H^\prime_P H^\pp_{A}}{N_1^\prime N_1^\pp N_2} S^{PA}_{\mu\nu}\,\vp^{\pp*\nu},
 \label{eq:BtoA}
 \end{eqnarray}
where
 \begin{eqnarray}
S^{PA}_{\mu\nu} &=&(S^{PA}_{V}-S^{PA}_A)_{\mu\nu}
  \nonumber\\
                &=&{\rm Tr}\left[\left(\gamma_\nu-\frac{1}{W^\pp_{A}}
                (p_1^\pp-p_2)_\nu\right)
                                 \gamma_5
                                 (\not \!p^\pp_1+m_1^\pp)
                                 (\gamma_\mu-\gamma_\mu\gamma_5)
                                 (\not \!p^\prime_1+m_1^\prime)\gamma_5(-\not
                                 \!p_2+m_2)\right]
 \nonumber\\
                &=&{\rm Tr}\left[\left(\gamma_\nu-\frac{1}{W^\pp_{A}}
                (p_1^\pp-p_2)_\nu\right)
                                 (\not \!p^\pp_1-m_1^\pp)
                                 (\gamma_\mu\gamma_5-\gamma_\mu)
                                 (\not \!p^\prime_1+m_1^\prime)\gamma_5(-\not
                                 \!p_2+m_2)\right].\label{eq:BtoAp}
 \end{eqnarray}
By comparing eq.~(\ref{eq:BtoV}) and eq.~(\ref{eq:BtoAp}), we have
$S^{PA}_{V(A)}=S^{PV}_{A(V)}$ with the replacement $m_1^\pp\to
-m_1^\pp,\,W^\pp_V\to W^\pp_{A}$, except for a phase $i$.
Consequently, the form factors of $B\to A$ can be related to the
$B\to V$ form factors through:
 \begin{eqnarray}
 \ell^{A}(q^2)&=&f(q^2) \,\,\,{\rm with}\,\,\,
                         (m_1^\pp\to -m_1^\pp,\,h^\pp_V\to h^\pp_{A},
                         \,w^\pp_V\to w^\pp_{A}),
 \nonumber\\
 q^{A}(q^2)&=&g(q^2) \,\,\,{\rm with}\,\,\,
                         (m_1^\pp\to -m_1^\pp,\,h^\pp_V\to h^\pp_{A},
                         \,w^\pp_V\to  w^\pp_{A}),
  \nonumber\\
 c_+^{A}(q^2)&=&a_+(q^2) \,\,\,{\rm with}\,\,\,
                          (m_1^\pp\to -m_1^\pp,\,h^\pp_V\to h^\pp_{A},
                          \,w^\pp_V\to  w^\pp_{A}),
                          \label{eq:BtoAformfactor}
 \end{eqnarray}
where we should be cautious that the replacement of $m_1^\pp\to
-m_1^\pp$ can not be applied to $m_1^\pp$ in $w^\pp$ and $h^\pp$.
 \begin{eqnarray}
 A^{PA}(q^2)&=&-(m_{B_c}-m_X)\, q(q^2),\quad
 V^{PA}_1(q^2)=-\frac{\ell(q^2)}{m_{B_c}-m_X},
 \nonumber\\
 V^{PA}_2(q^2)&=&(m_{B_c}-m_X)\, c_+(q^2).
 \end{eqnarray}

In the above expressions for the form factors, there are many
terms containing $(p_\perp^\prime\cdot q_\perp)/q^2$ in the
integrand. These terms can make non-trivial contributions together
with $h^{\prime\prime}_M/\hat N_1^{\prime\prime}$. In the
calculation, we make a Taylor expansion for
$h^{\prime\prime}_M/\hat N_1^{\prime\prime}$ as:
\begin{eqnarray}
 \frac{h^\pp_V}{\hat N_1^\pp}
 =\frac{h^\pp_V}{\hat  N_1^\pp}\Big |_{p^{\pp2}_\bot\to p^{\prime  2}_\bot}
 -2\, x_2 p^\prime_\bot\cdot q_\bot\Big(\frac{d}{dp^{\pp2}_\bot}\frac{h^\pp_V}
 {\hat N_1^\pp}\Big)_{p^{\pp2}_\bot\to p^{\prime
 2}_\bot}+{\cal O}(x^2_2q^2).
\end{eqnarray}
Then terms containing $p_\perp^\prime\cdot q_\perp$ can be
simplified using the following equation:
 \begin{eqnarray}
 \int d^2p'_\bot\frac{(p'_\bot\cdot q_\bot)^2}{q^2}=-{1\over 2}\int
 d^2p'_\bot\,p'^2_\bot.
 \end{eqnarray}

Now it is straightforward to obtain the decay width:
\begin{eqnarray}
\Gamma(B_c^-\to J/\psi\pi^-)&=&\frac{|G_FV_{cb}V_{ud}^*a_1f_\pi
m_{B_c}^2 A_0^{PV}(0)|^2}{32\pi m_{B_c}}(1-r_{J/\psi}^2),
\label{eq:decaywidth}
\end{eqnarray}
where $r_{J/\psi}=\frac{m_{J/\psi}}{m_{B_c}}$. For the decays
involving $K^-$, the factor $V_{ud}^*f_\pi$ is replaced by
$V_{us}^*f_K$; while for $B_c^-\to X\pi^-(K^-)$,
$A^{PV}_0(0)$($r_{J/\psi}$) is replaced by $V^{PA}_0(0)$
($r_{X}$).

\section{Numerical Results and Discussion}
\label{III}

 To perform numerical calculations we need to specify the
input parameters in the covariant light-front framework. The $\bar
qq$ meson state is described by the light-front wave function
which can be obtained by solving the relativistic Schr\"odinger
equation with a phenomenological potential. But in fact except for
some special cases, the solution is not obtainable at present. We
prefer to  employ a phenomenological wave function to describe the
hadronic structure. In the present work, we shall use the simple
Gaussian-type wave function \cite{Gauss}
\begin{eqnarray} \label{eq:Gauss}
 \varphi^\prime
    &=&\varphi^\prime(x_2,p^\prime_\perp)
             =4 \left({\pi\over{\beta^{\prime2}}}\right)^{3\over{4}}
               \sqrt{{dp^\prime_z\over{dx_2}}}~{\rm exp}
               \left(-{p^{\prime2}_z+p^{\prime2}_\bot\over{2 \beta^{\prime2}}}\right),
\nonumber\\
 \varphi^\prime_p
    &=&\varphi^\prime_p(x_2,p^\prime_\perp)=\sqrt{2\over{\beta^{\prime2}}}
    ~\varphi^\prime,\quad\qquad
         \frac{dp^\prime_z}{dx_2}=\frac{e^\prime_1 e_2}{x_1 x_2 M^\prime_0}.
 \label{eq:wavefn}
\end{eqnarray}
The input parameters $m_q$ and $\beta$ in the Gaussian-type wave
function (\ref{eq:wavefn}) are shown in Table~\ref{tab:input}. The
constituent quark masses are close to those used in the
literature~\cite{Jaus1,CCH1,Jaus2,CCH2,Cheng:2004yj,LFHW}. The
parameter $\beta'$, which describes the momentum distribution, is
expected to be of order $\Lambda_{\rm QCD}$. These parameters
$\beta$'s are fixed by the decay constants whose analytic
expressions in the covariant light-front model are given in
\cite{CCH2}. The decay constant $f_{J/\psi}$ can be determined by
the leptonic decay width:
\begin{eqnarray}
\Gamma_{ee}\equiv\Gamma(J/\psi\to e^+e^-)=\frac{4\pi
\alpha_{em}^2Q_c^2 f_{J/\psi}^2}{3m_{J/\psi}},
\end{eqnarray}
where $Q_c=2/3$, denotes the electric charge of the charm quark.
Using the measured results for the electronic width of
$J/\psi$~\cite{PDG}:
\begin{eqnarray}\Gamma_{ee}=(5.55\pm0.14\pm0.02)~\mbox{keV},
\end{eqnarray}
we obtain $f_{J/\psi}=0.416\pm 0.05$~GeV. As for the decay constant
for $B_c$ and $X$, we use $f_{B_c}=0.398^{+0.054}_{-0.055}$~GeV, and
$|f_{X(3872)}|=0.329^{+0.111}_{-0.095}$~GeV. For the light
pseudoscalars, we use $f_{\pi}=0.132$ GeV and $f_{K}=0.16$ GeV. The
determined results for $\beta$ are listed in Table~\ref{tab:input}.

\begin{table}
\caption{ The input parameters $m_q$ and $\beta$ (in unit of GeV)
in the Gaussian-type light-front wave function (\ref{eq:wavefn}).}
\begin{ruledtabular}
\begin{tabular}{cccccc}
          \  $m_c$
          & $m_b$
          & $\beta_{B_c}$
          & $\beta_{J/\psi}$
          & $\beta_{X}$
          \\
\hline     $1.4$
          & $4.4$
          & $0.870\pm0.100$
          & $0.631^{+0.06}_{-0.04}$
          & $0.720\pm0.100$
\end{tabular}\label{tab:input}
\end{ruledtabular}
\end{table}

In the factorization approach,  the decay amplitude is expressed as
a product of the short distance Wilson coefficients, the form
factors and meson decay constant. The latter two are physical values
which are scale-independent. But the Wilson coefficient $a_1$ of the
four quark operators depends on the factorization scale.  This
directly leads to the scale dependence of the decay amplitude. But
as we have shown in Ref.~\cite{PQCDBs}, the numerical value of $a_1$
is not very sensitive to the scale thus we use $a_1=1.1$ in this
work. The CKM matrix elements, lifetime of $B_c$ meson and the
masses of the hadrons are chosen from the Particle Data
Group~\cite{PDG} as:
\begin{eqnarray}
&&|V_{cb}|=0.0416\pm0.0006,\;\; |V_{ud}|=0.97377\pm0.00027,\;\;\; |V_{us}|=0.2257\pm0.021,\\
&&m_{B_c}=6.286~\mbox{GeV},\tau_{B_c}=(0.46^{+0.18}_{-0.16})\times
10^{-12}s,\\
&&m_{J/\psi}=3.097~\mbox{GeV},\;\; m_{X}=3.872~\mbox{GeV}.
\end{eqnarray}
The uncertainties in the above CKM matrix elements are small, thus
induce small errors to the decay width and we will neglect these
uncertainties.

Using the above inputs, we can calculate the form factors directly.
As in Refs.~\cite{Jaus2,CCH2}, because of the condition $q^+=0$ we
have imposed during the course of calculation, form factors are
known only for spacelike momentum transfer $q^2=-q^2_\bot\leq 0$.
But only the timelike form factors are relevant for the physical
decay processes. It has been proposed in \cite{Jaus1} to recast the
form factors as explicit functions of $q^2$ in the spacelike region
and then analytically extrapolate them to the timelike region. In
the exclusive non-leptonic decays, only the form factor at maximally
recoiling ($q^2\simeq 0$) is required therefore we do not need to
discuss the dependence on the momentum transfer here. After
calculation, the results for the $B_c \to J/\psi$ and $B_c \to X$
(assuming $X$ as a $1^{++}$ state) form factors are
\begin{eqnarray}
V^{PV}(0)=0.87^{+0.00+0.01}_{-0.02-0.00},&& A^{PV}_0(0)=A^{PV}_3(0)=0.57^{+0.01+0.00}_{-0.02-0.00},\nonumber\\
A^{PV}_1(0)=0.55^{+0.01+0.00}_{-0.03-0.00},&&  A^{PV}_2(0)=0.51^{+0.03+0.00}_{-0.04-0.00},\nonumber\\
A^{PA}(0)=0.36^{+0.02+0.01}_{-0.02-0.03},&&  V^{PA}_0(0)=V^{PA}_3(0)=0.18^{+0.01+0.01}_{-0.02-0.02},\nonumber\\
V^{PA}_1(0)=1.15^{+0.03+0.03}_{-0.04-0.06},&&
V^{PA}_2(0)=0.13^{+0.02+0.00}_{-0.02-0.01},\label{eq:formfactorresults}
\end{eqnarray}
where the first uncertainty is from the decay constant of the $B_c$
meson and the latter is from the charmonium decay constant. In
Table~\ref{Tab:formfactor}, we make a comparison of our results with
the previous studies. We can see that our results are slightly
smaller than the results from the three points sum rule and quark
model, but the light-cone sum rule predictions are quite different
 from the others.

\begin{table}\caption{The values of the form factors of $B_c\to J/\psi$ at $q^2 = 0$
in comparison with the estimates in the three points sum rule
(3PSR) (with the Coloumb corrections included)~\cite{3PSR}, in the
quark model (QM)~\cite{QM} and the light-cone sum rules
(LCSR)~\cite{LCSR} }\label{Tab:formfactor}
\begin{center}
\begin{tabular}{c|c|c|c}
\hline \hline
                           & $A_1$       &$A_2$   &$V$ \\
 \hline
3PSR~\cite{3PSR}            &$0.63$        &$0.69$      &$ 1.03$        \\
QM~\cite{QM}              & $0.68$     &$0.66$      &$0.96$            \\
LCSR~\cite{LCSR}            & $0.75$     &$1.69$      &$1.69$            \\
This work       & $0.55^{+0.01+0.00}_{-0.03-0.00}$     &$0.51^{+0.03+0.00}_{-0.04-0.00}$      &$0.87^{+0.00+0.01}_{-0.02-0.00}$           \\
 \hline \hline
\end{tabular}
\end{center}
\end{table}

Using the results for $B_c\to J/\psi$ form factors, we obtain the
branching ratio of $B_c^-\to J/\psi\pi^-(K^-)$:
\begin{eqnarray}
{\cal BR}(B_c^-\to J/\psi\pi^-)&=&(2.0^{+0.8+0.0+0.0}_{-0.7-0.1-0.0})\times 10^{-3},\nonumber\\
{\cal BR}(B_c^-\to J/\psi
K^-)&=&(1.6^{+0.6+0.0+0.0}_{-0.6-0.1-0.0})\times 10^{-4},
\end{eqnarray}
where the uncertainties are from the large errors in the lifetime of
$B_c$ meson, the decay constant of $B_c$ meson and the charmonium
decay constants. In the literature, these decays have received
extensively study \cite{BCtoJpi} and the range of the branching
ratios is
\begin{eqnarray}
{\cal BR}(B_c^-\to J/\psi \pi^-)&=&(0.06-0.18)\%,\nonumber\\
{\cal BR}(B_c^-\to J/\psi K^-)&=&(0.005-0.014)\%,
\end{eqnarray}
which are consistent with ours. Assuming $X$ as a $1^{++}$ state,
the branching ratios of $B_c^-\to X(3872)\pi^-(K^-)$ are
\begin{eqnarray}
{\cal BR}(B_c^-\to X(3872)\pi^-)&=&(1.7^{+0.7+0.1+0.4}_{-0.6-0.2-0.4})\times 10^{-4},\nonumber\\
{\cal BR}(B_c^-\to
X(3872)K^-)&=&(1.3^{+0.5+0.1+0.3}_{-0.5-0.2-0.3})\times
10^{-5}.\label{eq:BRBtoX1}
\end{eqnarray}
These results are one order magnitude smaller than the branching
ratio of $B_c^-\to J/\psi\pi^-$ and $B_c^-\to J/\psi K^-$,
respectively. From the decay width formulae in
Eq.~(\ref{eq:decaywidth}), we know that the $B_c\to J/\psi P$
branching ratios are proportional to the form factor $|A^{PV}(0)|^2$
while for $B_c\to X(1^{++})P$, the decay width are proportional to
$|V^{PA}(0)|^2$. The $B_c\to X$ form factor $V^{PA}(0)$ is only
$1/3$ of $|A^{PV}(0)|$ shown in Eq.~(\ref{eq:formfactorresults}),
which induces the one order magnitude difference for these two kinds
of decays.

For $B_c\to VV$ decays, there are three different polarizations.
According to the power counting rule in the standard
model~\cite{powercounting}, the longitudinal polarization dominates
in the decay processes, while other polarizations suffer one or two
orders of $\Lambda_{QCD}/m_B$ or $m_c/m_B$ suppressions which arise
 from the quark helicity flip. It is found that the annihilation
diagrams with the operator $O_6$ could violate this power counting
rule~\cite{PQCD}. However in the $B_c^-\to J/\psi\rho^-(K^{*-})$
decays, there are only emission type contributions thus the power
counting rule should work well. If we neglect the
$m_\rho^2/m_{B_c}^2$ terms in the polarization vector, the formulae
for the branching ratios of $B\to VV$ is the same as
eq.~(\ref{eq:decaywidth}) with the replacement of the decay
constant: $f_P\to f_V$. Using $f_\rho=0.209$~GeV and
$f_{K^*}=0.217$~GeV, we obtain the corresponding branching ratios
as:
\begin{eqnarray}
{\cal BR}(B_c^-\to J/\psi\rho^-)&=&(5.0^{+2.0+0.1+0.1}_{-1.7-0.0-0.0})\times 10^{-3},\nonumber\\
{\cal BR}(B_c^-\to J/\psi
K^{*-})&=&(2.9^{+1.1+0.0+0.0}_{-1.0-0.0-0.0})\times
10^{-4},\nonumber\\
{\cal BR}(B_c^-\to X(3872)\rho^-)&=&(4.1^{+1.6+0.3+0.1}_{-1.4-0.1-0.1})\times 10^{-4},\nonumber\\
{\cal BR}(B_c^-\to
X(3872)K^{*-})&=&(2.4^{+0.9+0.2+0.5}_{-0.8-0.3-0.5})\times
10^{-5}.\label{eq:BRBtoX2}
\end{eqnarray}

Our above calculation is based on the $1^{++}$ charmonium
description for $X$.  The charmonium states with other quantum
numbers can also be studied similarly in this approach. If the
quantum numbers of $X(3872)$ are changed to $1^{--}$, the large
form factor $A_0^{B_c\to X}$ can enhance the production rates
dramatically as:
\begin{eqnarray}
 {\cal BR}(B_c^-\to X(3872)\pi^-)&=& (1.4^{+0.6+0.0+0.4}_{-0.5-0.0-0.5}) \times 10^{-3},\nonumber\\
 {\cal BR}(B_c^-\to X(3872)K^{-})&=&(1.1^{+0.4+0.0+0.3}_{-0.4-0.0-0.4}) \times 10^{-4},\nonumber\\
 {\cal BR}(B_c^-\to X(3872)\rho^-)&=&(3.5^{+1.4+0.0+1.1}_{-1.2-0.2-1.2}) \times 10^{-3},\nonumber\\
 {\cal BR}(B_c^-\to X(3872)K^{*-})&=&(2.0^{+0.8+0.0+0.6}_{-0.7-0.1-0.7})
 \times10^{-4}.\label{eq:BRBtoXvector}
\end{eqnarray}
Comparing the above equations with eqs.~(\ref{eq:BRBtoX1}) and
(\ref{eq:BRBtoX2}), we can see that the production rates are
enhanced by about one order magnitude.  The large branching ratios
and large difference between different quantum number of state X can
  be easily used  at LHCb experiment to test the charmonium description for $X$.

\section{Conclusion}
\label{conc}

In the covariant light-front quark model, we study the form factors
of $B_c\to J/\psi$ and $B_c\to X$ transitions at maximum recoiling.
The factorization of exclusive processes $B_c^- \to
J/\psi\pi^-(K^-)$ and $B_c^- \to X(3872)\pi^-(K^-)$ can be proved to
all orders of the strong coupling constant just as the proof in
$\bar B^0\to D^+\pi^-$ and $B^-\to D^0\pi^-$. Therefore the decay
width of these decays can be simply calculated in the naive
factorization approach utilizing the form factors. Our results for
the branching ratio of $B_c\to J/\psi \pi^-(K^-)$ are consistent
with the previous studies considering the uncertainties. The study
of these exclusive processes may greatly improve our understanding
on the $B_c$ meson exclusive hadronic decays, and the corresponding
theory describing them as well.  In our calculation, identifying
$X(3872)$ as a $1^{++}$ charmonium state, we obtain ${\cal
BR}(B_c^-\to X(3872)\pi^-)=(1.7^{+0.7+0.1+0.4}_{-0.6-0.2-0.4})\times
10^{-4}$ and ${\cal BR}(B_c^-\to
X(3872)K^-)=(1.3^{+0.5+0.1+0.3}_{-0.5-0.2-0.3})\times 10^{-5}$. If
assuming $X$ as a $1^{--}$ state, the branching ratios are one order
magnitude larger.  This large difference can be easily used by the
 LHCb experiment to test the different charmonium description for $X$.

\section*{Acknowledgement}

We thank H.Y. Cheng, C.K. Chua and Y.M. Wang for useful
discussions. This work is partly supported by National Science
Foundation of China under Grant No.~10475085 and 10625525.


\end{document}